\documentclass[prb,twocolumn,amsmath,amssymb,showpacs]{revtex4}

\usepackage{graphicx}
\usepackage{dcolumn}
\usepackage{bm}
\usepackage{mathrsfs}
\usepackage{appendix}
\usepackage{bbm}
\usepackage{color}

\def \vS {{\bf S}}
\def \vR {{\bf R}}
\def \ve {{\bf e}}
\def \vQ {{\bf Q}}
\def \vq {{\bf q}}

\def \ve {{\bf e}}
\def \vm {{\bf m}}

\def \mb {\mu_{\rm B}}

\def \xp {{\bf x}^{\prime }}
\def \yp {{\bf y}^{\prime }}
\def \zp {{\bf z}^{\prime }}
\def \xx {{\bf x}}
\def \yy {{\bf y}}
\def \zz {{\bf z}}

\def \vt {\boldsymbol{\tau}}
\def \vM {{\bf M}}
\def \vP {{\bf P}}
\def \BF {{\rm BiFeO$_3$} }
\def \BP {{\rm BiFeO$_3$}}

\def \TN {T_{\rm N}}

\def \XX {{\bf X}}
\def \YY {{\bf Y}}
\def \ZZ {{\bf Z}}

\def \vn {{\bf n}}

\def \vB {{\bf B}}

\def \Bp {B_{\rm pin}}

\def \hi {{h_i}}

\def \tpin {\tau_{{\rm pin}}}
\def \chp {\chi_{\perp }}
\def \cpa {\chi_{\parallel}}
\def \epin {\varepsilon_{\rm pin}}
\def \eps {\varepsilon }

\begin{document}

\title{Pinning, Rotation, and Metastability of BiFeO$_3$ Cycloidal Domains in a Magnetic Field
\footnote{Copyright notice: This manuscript has been authored by UT-Battelle, LLC under Contract No. DE-AC05-00OR22725 with the U.S. Department of Energy. The United States Government retains and the publisher, by accepting the article for publication, acknowledges that the United States Government retains a non-exclusive, paid-up, irrevocable, world-wide license to publish or reproduce the published form of this manuscript, or allow others to do so, for United States Government purposes. The Department of Energy will provide public access to these results of federally sponsored research in accordance with the DOE Public Access Plan (http://energy.gov/downloads/doe-public-access-plan).}}

\author{Randy S. Fishman}

\affiliation{Materials Science and Technology Division, Oak Ridge National Laboratory, Oak Ridge, Tennessee 37831, USA}

\date{\today}

\begin{abstract}

Earlier models for the room-temperature multiferroic \BF implicitly assumed that a very strong anisotropy restricts the domain wavevectors $\vq $ to the three-fold symmetric axis normal
to the static polarization $\vP $.  However, recent measurements demonstrate that the domain wavevectors rotate so that $\vq $ rotates
within the hexagonal plane normal to $\vP $ away from the field orientation $\vm$.  We show that the previously neglected three-fold anisotropy $K_3$ restricts
the wavevectors to lie along the three-fold axis in zero field.  For $\vm$ along a three-fold axis, the domain with $\vq $ parallel to $\vm $ remains metastable 
below $B_{c1}\approx 7$ T.  Due to the pinning of domains by non-magnetic impurities, the wavevectors of the other two domains start to rotate away 
from $\vm $ above 5.6 T, when the component of the torque $\vt = \vM \times \vB $ along $\vP $ exceeds a threshold value $\tpin $.
Since $\vt =0$ when $\vm \perp \vq $, the wavevectors of those domains never become completely perpendicular to the magnetic field.  
Our results explain recent measurements of the critical field as a function of field orientation, small-angle neutron scattering 
measurements of the wavevectors, as well as spectroscopic measurements with $\vm $ along a three-fold axis.

\end{abstract}


\maketitle

\section{Introduction}

The manipulation of magnetic domains with electric and magnetic fields is one of the central themes [\onlinecite{eer06, zhao06, tokunaga15}] 
in the study of multiferroic materials.  Applications of multiferroic materials depend on a detailed understanding of how domains respond to 
external probes.  Despite recent advances [\onlinecite{park14}] in our understanding of the room-temperature multiferroic \BP, 
however, some crucial questions remain about how its cycloidal domains respond to a magnetic field.  

A type I or ``proper" multiferroic, \BF exhibits 
a strong ferroelectric polarization of about 80 $\mu $C/cm$^2$ along one of the pseudo-cubic diagonals
below the ferroelectric transition at $T_{FE} = 1100$ K [\onlinecite{teague70, lebeugle07}]. 
Below $T_{FE}$, broken symmetry produces two Dzaloshinskii-Moriya (DM) interactions 
between the $S=5/2$ Fe$^{3+}$ ions.  A magnetic transition at $\TN = 640$ K [\onlinecite{sosnowska82}] 
allows the cycloidal spin state to take advantage of this broken symmetry.

Until recently, it seemed that a complete theoretical description [\onlinecite{sosnowska95, sousa08, pyat09, rahmedov12, fishman13a, fishman13c}]
of rhombohedral \BF was in hand.  Employing the first available single crystals, the measured cycloidal 
frequencies [\onlinecite{rovillain09,talb11,nagel13}] of \BF provided a stringent test for theory.  A microscopic model for \BF that includes two 
DM interactions $D_1$ and $D_2$ and single-ion anisotropy $K_1$ successfully predicted [\onlinecite{fishman13a, fishman13c}]
the mode frequencies in zero field [\onlinecite{rovillain09,talb11}] and their magnetic field evolution [\onlinecite{nagel13}] for several field orientations.  
Since all model parameters were determined [\onlinecite{param}] from the zero-field behavior of \BP, the field evolution of the cycloidal modes [\onlinecite{fishman13c}] 
provided a particularly good test of the microscopic model.  Nevertheless, new evidence suggests that this model is not complete.

With the electric polarization $\vP = P\zp $ along the pseudo-cubic diagonal $[1,1,1]$ ($[a,b,c]$ is a unit vector), 
the three magnetic domains of \BF in zero field [\onlinecite{rama11}] have 
wavevectors $\vQ_k = \vQ_0 + \vq_k$ where $\vQ_0= (\pi /a)(1,1,1)$ is the antiferromagnetic (AF) Bragg vector,
\begin{eqnarray}
\vq_1 &=& \frac{2\pi \delta }{a}(-1, 1 ,0 ), \\
\vq_2 &=& \frac{2\pi \delta }{a} (1 ,0 -1), \\
\vq_3 &=& \frac{2\pi \delta }{a}(0,-1 ,1 ),
\end{eqnarray}
$a = 3.96$ \AA $ $ is the lattice constant of the pseudo-cubic unit cell, and
$\delta \approx 0.0045$ determines the cycloidal wavelength $\lambda = a/\sqrt{2}\delta \approx 620$ \AA .
In zero field, the three domains of \BF with wavevectors $\vq_k$ are degenerate.  For each domain $k$, the spins of the cycloid 
lie primarily in the plane defined by $\zp = [1,1,1]$ and $\xp $, which is the unit vector along $\vq_k$.   A magnetic field favors
domains with $\xp \perp \vB $ because $\chp \gg \cpa $ [\onlinecite{leb07}] for \BP. 

\begin{figure}
\includegraphics[width=8.3cm]{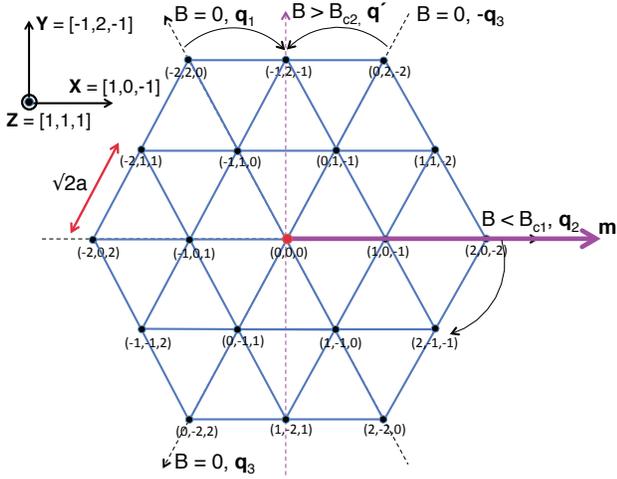}
\caption{(Color online) A hexagonal plane normal to $\ZZ $ with $\vm = \XX $.  In zero field, three domains with
wavevectors $\vq_k$ are stable.  Wavevectors $\vq_1$ and $\vq_3$ rotate towards $\YY \perp \vm $ with increasing field.  
A domain with wavevector $\vq' \parallel \YY $ is stable above $B_{c2}$ in the absence of pinning.  
Below $B_{c1}$, the domain with wavevector $\vq_2$ is metastable.}
\end{figure}

The model described above successfully predicted the field dependence of the mode frequencies [\onlinecite{nagel13,fishman13c}] 
when the stable domain has wavevector $\vq \equiv \vQ -\vQ_0$ perpendicular to the magnetic field $\vB = B\vm $.
For example, the mode frequencies with $\vm = [0,0,1]$ and domain 1 or with $\vm = [0,1,1]$ and domain 3 almost exactly match the theoretical predictions. 
But for $\vm $ along a three-fold symmetric axis like $[1,0,-1]$, the mode frequencies are not as well described by taking the 
domain wavevector $\vq $ along $[-1,1,0]$ or $[0,-1,1]$.  
Rather, the mode frequencies are then virtually identical to those with $\vm = [-1,2,-1]$ and $\vq $ along
$[1,0,-1] \perp \vm$ [\onlinecite{farkasun}].  In addition, the selection rules for the appearance of the spectroscopic modes
do not follow the expected rules when the domain wavevectors lie along the three-fold axis [\onlinecite{mats16}].
So it appears that for a magnetic field along $\vm = [1,0,-1]$,
the domain wavectors $\vq_1$ and $\vq_3$ rotate away from $\vm $ towards $[-1,2,-1]$, as indicated in Fig.1.

Another discrepancy appears in measurements of the critical field $B_{c3}(\vm )$ above which the canted AF (CAF) phase becomes stable.
Predictions based on the ``canonical" model indicate that $B_{c3}(\vm )$ depends on the stable domain as $\vm $ is rotated about $\zp = [1,1,1]$ by the azimuthal angle 
$\zeta $ [\onlinecite{fishman13b}].  However, experimental measurements [\onlinecite{tokunaga10, park11}] find that $B_{c3}(\vm )$ depends primarily on the polar angle 
$\vartheta = \cos^{-1} (\vm \cdot \zp )$ and does not sensitively depend on the azimuthal angle $\zeta $.

Direct evidence for domain rotation in a magnetic field was recently provided by small angle neutron-scattering (SANS)
[\onlinecite{bordacsun}].  Those measurements indicate that the metastable domain with wavevector $\vq  $ along
$\vm $ is slowly depopulated with increasing field and disappears above about 7 T.  The other two domains 
start to merge into a single domain with wavevector perpendicular to $\vm $ but $\vq \cdot \vm $ never reaches zero.

This behavior is caused by the pinning of domains by non-magnetic impurities.  In a ferromagnet [\onlinecite{lemerle98, kleemann07}], 
domain walls move when the component of $\vB $ along the domain magnetization $\vM$ exceeds the pinning field $\Bp $.   
In the strong-pinning limit with $\vm $ along a three-fold axis, 
cycloidal wavevectors $\vq $ begin to rotate away from $\vm $ when the component of the 
torque $\vt = \vM \times \vB $ along $\zp $ exceeds a threshold value $\tpin $.  Since $\vM $ is induced by the 
component of $\vB $ perpendicular to the cycloidal plane containing $\vq $, $\vt =0$ when 
$\vq \perp \vm $.  Consequently, the wavevector $\vq $
never becomes completely perpendicular to the external field unless it lies along a three-fold axis perpendicular to $\vm $.

This paper improves the ``canonical" model of \BF to address the discrepancies described above.  
Section II discuss the present ``canonical" model for \BP.
In Section III, we present the higher-order anisotropy terms that break three-fold symmetry.  The next two sections describe
the consequences of this modified model in the absence of pinning.
Section IV.A treats the case where $\vm $ lies along a three-fold axis so that the wavevectors of the stable domains rotate away from
the other two three-fold axis with increasing field.  Section IV.B treats the case where $\vm $ is perpendicular to a three-fold axis so that the 
wavevector of the stable domain does not rotate.  Section V discusses the effects of pinning and provides an exact solution for 
domain rotation in the strong-pinning limit.  We modify
the conclusions of Section IV to include the effects of pinning in Section VI.  Section VII contains a conclusion.  
The magnetoelastic coupling of \BF is examined in the Appendix.

\section{The ``Canonical" Model}

The ``canonical" model of \BF has the Hamiltonian
\begin{eqnarray}
\label{Ham}
&&{\cal H} = -J_1\sum_{\langle i,j\rangle }\vS_i\cdot \vS_j -J_2\sum_{\langle i,j \rangle'} \vS_i\cdot \vS_j
\nonumber \\
&&+D_1\, \sum_{\langle i,j\rangle } (\zp \times {\bf e}_{i,j}/a) \cdot (\vS_i\times\vS_j) \nonumber \\
&& + D_2 \, \sum_{\langle i,j\rangle } \, (-1)^\hi \,\zp \cdot  (\vS_i\times\vS_j)\nonumber \\
&& -K_1\sum_i (\zp \cdot \vS_i )^2
- 2\mb B \sum_i \vm \cdot \vS_i ,
\end{eqnarray}
where $\ve_{i,j } = a{\bf x}$, $a{\bf y}$, or $a{\bf z}$ connects the spin $\vS_i$ on site $\vR_i$ with its nearest neighbor $\vS_j$ 
on site $\vR_j=\vR_i + \ve_{i,j}$.  The integer layer number
$h_i$ is defined by $\sqrt{3} \vR_i \cdot \zp /a$.
The first DM interaction $D_1$ determines the cycloidal wavelength $\lambda $;  
the second DM interaction $D_2$ produces a small tilt $\tau \approx 0.3^o$ of the spins out of the $\xp - \zp$ plane [\onlinecite{sosnowska95, pyat09}].   

\begin{table}
\caption{Reference frames of BiFeO$_3$}
\begin{ruledtabular}
\begin{tabular}{cc}
unit vectors & description and values\\
\hline
$\{\xx, \yy  ,\zz \}$ & pseudo-cubic unit vectors  \\
& $\xx = [1,0,0]$, $\yy = [0,1,0]$, $\zz = [0,0,1]$ \\
$\{\xp ,\yp ,\zp \}$ & rotating reference frame of cycloid \\
&  $\xp \parallel \vq $, $\zp = [1,1,1]$, $\yp = \zp \times \xp $ \\
$\{ \XX , \YY , \ZZ \}$ & fixed reference frame of hexagonal plane \\
& $\XX = [1,0,-1]$, $\YY = [-1,2,-1]$, $\ZZ = [1,1,1]$\\
\end{tabular}
\end{ruledtabular}
\end{table}

The first DM term in ${\cal H}$,
\begin{equation}
{\cal H}_{D_1} = D_1\, \sum_{\langle i,j\rangle } (\zp \times {\bf e}_{i,j}/a) \cdot (\vS_i\times\vS_j),
\end{equation}
does not depend on the choice of domain and $\vq_k$.
In earlier [\onlinecite{park14}] versions of the ``canonical" model, this term was restricted to a specific domain of the cycloid.
For domain 2 with $\vq_2 $ parallel to $\xp = [1,0,-1]$ and $\yp = \zp \times \xp = [-1,2,-1]$, it was written
\begin{equation}
{\cal H}_{D_1}' =  -\frac{D_1}{\sqrt{2}}\sum_{\vR_j=\vR_i + a(\xx -\zz )} \,\yp \cdot (\vS _i\times \vS_j), 
\end{equation}
where $\vR_i$ and $\vR_j$ are next-nearest neighbors of the pseudo-cubic unit cell that lie on the same hexagonal layer $h_i$.  
Because the wavelength of the cycloid is so long, ${\cal H'}$ has the same static and dynamical properties as ${\cal H}$
provided that ${\cal H}$ is applied to the domain specified by ${\cal H}_{D_1}'$.

Why replace ${\cal H'}$ with ${\cal H}$?  Unlike ${\cal H'}$, ${\cal H}$ can be used to study any domain
with $\vq_k$ along a three-fold axis.
As shown below, ${\cal H}$ also describes the general case where $\vq $ differs from a three-fold axis.  
While ${\cal H}_{D_1}'$ involves the sum over next-nearest neighbors, ${\cal H}_{D_1}$ involves the sum
over nearest neighbors, which should dominate the DM interaction.  
Finally, the general form of ${\cal H}_{D_1}$ given above was obtained from first-principles calculations [\onlinecite{lee16a}].

\begin{figure}
\includegraphics[width=7.6cm]{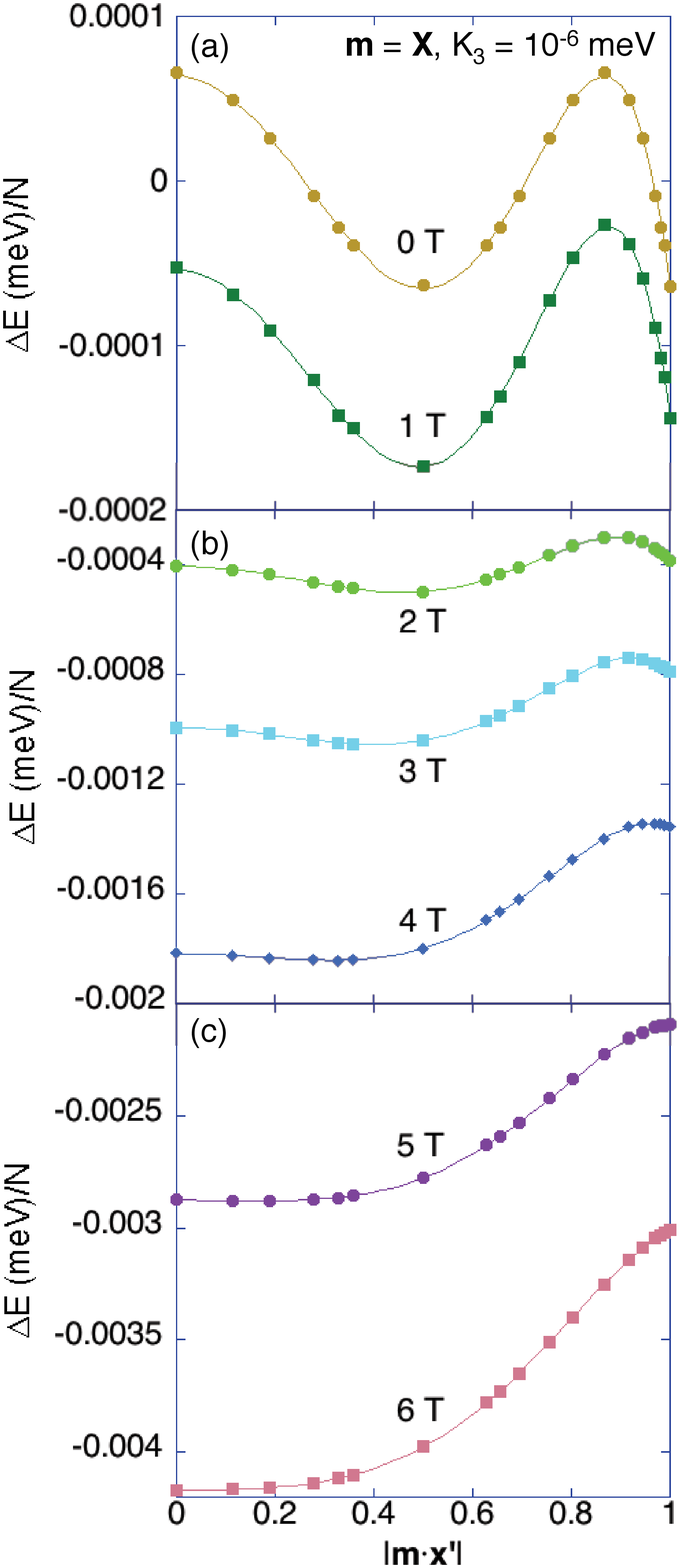}
\caption{(Color online) The energy $\Delta E/N $ versus wavevector $\vert \vm \cdot \xp \vert $ for $K_3 = 10^{-6}$ meV and field ranging from 0 to 6 T
along $\XX $.  Solid curves are fits to a sixth order polynomial in $\vert \vm \cdot \xp \vert $}
\end{figure}

To construct the local reference frame of a cycloid with wavevector $\vq = \vQ -\vQ_0$, we take
\begin{equation}
\vq = \frac{2\sqrt{2} \pi \delta }{a \vert \vn \vert } (n_x,n_y,n_z) =  \frac{2\pi }{\lambda }\xp ,
\end{equation}
where $n_i$ are integers with no common factors.  Then the unit vector along $\vq $ is $\xp = (n_x,n_y,n_z)/\vert \vn \vert $ and
$\yp = \zp \times \xp = (n_z-n_y, n_x-n_z, n_y-n_x)/(\sqrt{3} \vert \vn \vert )$.
With the local reference frame of a cycloid defined by the unit vectors $\{ \xp, \yp, \zp \}$,
the spin at site $\vR_i = (l,m,o)a$ is indexed by the integer 
$r = \vn \cdot \vR_i /a = n_x l + n_y m +n_z o$.  Assuming that the spins $\vS^{(j)}_r$ on alternate layers $j=1$ or 2 are identical functions of $r$, then
$r$ ranges from 1 to $M= \vert \vn \vert /\sqrt{2} \delta = \lambda \vert \vn \vert /a$ in the magnetic unit cell.

It is straightforward to show that 
\begin{eqnarray}
&&\frac{1}{N}{\cal H}_{D_1} = \frac{D_1}{2\sqrt{3} M} \sum_{r=1}^M \biggl\{ \\ &&
\,\,\,\,\, \xx \cdot \Bigl[ \vS^{(1)}_r \times \bigl( 
\vS^{(2)}_{r+n_z}-\vS^{(2)}_{r-n_z}  -\vS^{(2)}_{r+n_y}+\vS^{(2)}_{r+n_y}\bigr) \Bigr] \nonumber \\ &&
+\yy \cdot \Bigl[ \vS^{(1)}_r \times \bigl( 
\vS^{(2)}_{r+n_x}-\vS^{(2)}_{r-n_x}
-\vS^{(2)}_{r+n_z}+\vS^{(2)}_{r+n_z}\bigr) \Bigr] \nonumber \\ &&
+\zz \cdot \Bigl[ \vS^{(1)}_r \times \bigl( 
\vS^{(2)}_{r+n_y}-\vS^{(2)}_{r-n_y}
-\vS^{(2)}_{r+n_x}+\vS^{(2)}_{r+n_x}\bigr) \Bigr]\biggr\}.\nonumber 
\end{eqnarray}
Since $\vS^{(j)}_{r+M} =\vS^{(j)}_r$,  the index $r+n_{\alpha }$ can be taken mod $M$ to lie between 1 and $M$.
Because $\lambda /a = M/\vert \vn\vert \gg1$, 
$\vert n_i \vert \ll M$ and
\begin{equation}
\vS^{(2)}_{r+n_i }-\vS^{(2)}_{r-n_i} \approx n_i \bigl( \vS^{(2)}_{r+1}-\vS^{(2)}_{r-1} \bigr),
\end{equation}
with corrections of order $\delta^3 \sim 10^{-7}$.
This leads to the simpler form [\onlinecite{cont}]
\begin{equation}
\frac{1}{N}{\cal H}_{D_1} = \frac{D_1\vert \vn \vert }{2 M} \yp \cdot \sum_{r=1}^M \Bigl\{ \vS^{(1)}_r \times \bigl( 
\vS^{(2)}_{r+1}-\vS^{(2)}_{r-1} \bigr)\Bigr\}.
\end{equation}
Hence, the first DM interaction produces a cycloid in the
$\xp - \zp $ plane for any wavevector $\vq \parallel \xp $.

The second DM interaction, can be similarly written as [\onlinecite{lee16a}]
\begin{equation}
\frac{1}{N}{\cal H}_{D_2} = \frac{3D_2}{M} \zp \cdot \sum_{r=1}^M \bigl( \vS^{(1)}_r \times \vS^{(2)}_r \bigr),
\end{equation}
which rotates alternate layers of spins about the $\zp $ axis and tilts the cyloid out of the $\xp - \zp $ plane 

Neither of these DM interactions fixes the orientation of $\vq $ along a three-fold axis in zero field!
To remedy that deficiency, we must add an additional term to the Hamitlonian that breaks the three-fold symmetry
in the hexagonal plane perpendicular to $\zp $.

\section{Anisotropy energies}

Because $\{ \xp ,\yp ,\zp \}$ already provides the reference frame for the cycloid, which can rotate in the plane
perpendicular to $\zp $, we define $\XX = [1,0,-1]$ and $\YY = [-1,2,-1]$ as fixed axis in the hexagonal
plane.  Of course, $\ZZ = \XX \times \YY = [1,1,1]$ coincides with $\zp $ and lies along $\vP $. 
The three reference frames used in this paper are summarized in Table I.

 The lowest-order anisotropy energy of \BF was included in the 
``canonical" model:
\begin{equation}
{\cal H}_{K_1}= - K_1 \sum_i S_{iZ}^2.
\end{equation}
The two next-order anisotropy terms consistent with the rhombohedral symmetry [\onlinecite{weingart12}] of \BF are 
\begin{eqnarray}
{\cal H}_{K_2}&=& -\frac{1}{2} K_2\sum_i S_{iZ} \nonumber \\
&&\times \Bigl\{ \bigl(S_{iX}+iS_{iY}\bigr)^3 
+\bigl(S_{iX}-iS_{iY}\bigr)^3\Bigr\},
\end{eqnarray}
\begin{equation}
\label{xan} 
{\cal H}_{K_3}= -\frac{1}{2}K_3 \sum_i \Bigl\{ \bigl(S_{iX}+iS_{iY}\bigr)^6 + 
\bigl(S_{iX}-iS_{iY}\bigr)^6\Bigr\}.
\end{equation}
Whereas $K_1$ is of order $l^2\,\vert J_1\vert $ in terms of the dimensionless spin-orbit coupling constant $l$, 
$K_2$ and $K_3$ are of order $l^3\, \vert J_1\vert $ and $l^4\,\vert J_1\vert $, respectively [\onlinecite{bruno89}].
These three terms have classical energies
\begin{equation}
E_{K_1}= \langle {\cal H}_{K_1}\rangle = -S^2 K_1 \sum_i \cos^2 \theta_i,
\end{equation}
\begin{equation}
E_{K_2}= \langle {\cal H}_{K_2}\rangle = -S^4 K_2 \sum_i \cos\theta_i \,\sin^3 \theta_i \,\cos 3\phi_i,
\end{equation}
\begin{equation}
E_{K_3}= \langle {\cal H}_{K_3}\rangle = -S^6 K_3 \sum_i \sin^6 \theta_i \,\cos 6\phi_i,
\end{equation}
where the spin
\begin{equation}
\langle \vS_i \rangle = S\Bigl\{ \cos\phi_i \sin\theta_i \,\XX + \sin\phi_i \sin \theta_i \,\YY + \cos \theta_i \,\ZZ \Bigr\}
\end{equation}
is given in the fixed reference frame defined above.
Other anisotropy energies $S^2 K_1^{\prime } \sum_i \sin^2 \theta_i \,\cos 2\phi_i $ and $S^4 K_2^{\prime }\sum_i \sin^4 \theta_i \,\cos 4\phi_i $ 
vanish for the $R$3$c$ crystal structure of \BF [\onlinecite{weingart12, other}].

For the distorted cycloid of the ``canonical" model, both $E_{K_1}$ and $E_{K_3}$ are nonzero.  Because the cycloid is mirror symmetric about $Z=0$,
the summation in $E_{K_2}$ vanishes.
Therefore, $E_{K_2}$ will distort the cycloid to produce an energy reduction of order $-(K_2)^2/\vert J_1\vert $.  Since
$E_{K_2}/E_{K_3} \sim l^2 \ll 1$, $E_{K_2}$ can be neglected as a source of three-fold symmetry breaking compared to $E_{K_3}$.

\section{Magnetic fields}

In this section and the next, we neglect the effects of domain pinning.  The behavior of the domain wavevectors in an external field is then
completely determined by the model developed above.  The effects of pinning will be examined in Section V.

\begin{figure}
\includegraphics[width=8cm]{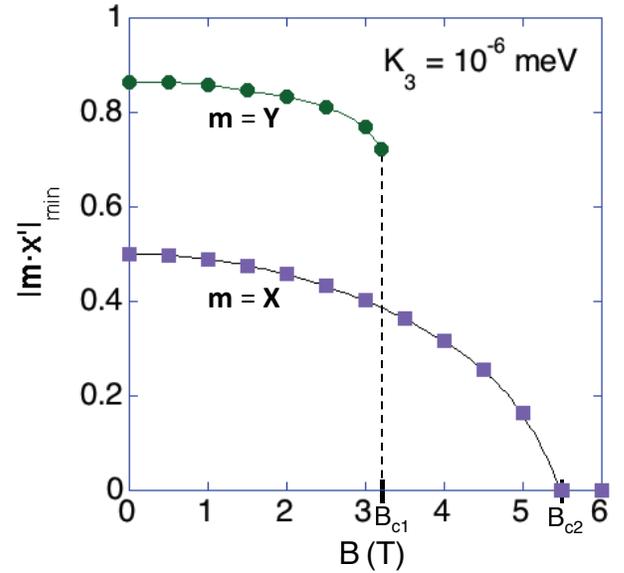}
\caption{(Color online) The wavevector $\vert \vm \cdot \xp \vert_{\rm min}$ for stable ($\vm = \XX $, squares) and metastable ($\vm = \YY $, circles)
domains versus field with $K_3 = 10^{-6}$ meV.}
\end{figure}

For $K_3 > 0$, $E_{K_3}$ favors spins that lie along one of the three three-fold axis $\phi_i =0$ and $\pm 2\pi/3$.
With this additional anisotropy, the wavevectors $\vq_k$ rotate away from the three-fold axis with increasing field when the field does not itself lie
perpendicular to a three-fold axis.

\subsection{Field along a three-fold axis or $\vm = \XX $}

First take the field along a three-fold axis such as $\XX $ in Fig.1.  
Assuming that the system has been cooled from high temperature in zero field,
all three domains with wavevectors $\vq_k$ will be equally occupied.  But in large field, we expect that the stable domain will
have wavevector $\vq' $ parallel to $\YY $ and perpendicular to $\vm $.  For $K_3= 10^{-6}$ meV, the energy $E = \langle \cal{H}\rangle $ is evaluated
for several different wavevectors at each field.  Defining $E_0$ as the energy for $K_3=0$ and $B=0$, 
results for $\Delta E = E- E_0$ are presented in Fig.2.

At zero field, $\Delta E$ is minimized when $\xp $ lies along a three-fold axis.  Since $\vm =\XX $ is itself a three-fold axis, 
minima appear when $\vert \vm \cdot \xp \vert = 1$ or $1/2$.  With increasing field, the minimum at $\vert  \vm \cdot \xp \vert = 1$ 
($\xp \parallel \vm$) increases in energy so that this solution is only metastable.  
The stable solutions rotate from $\vert  \vm \cdot \xp \vert  = 1/2$ towards 0.  

In addition to the critical field marking the transition into the CAF phase,
we identify two lower critical fields.  Below $B_{c1}\approx 4.6$ T, the minima at $\vert  \vm \cdot \xp \vert  = 1$ survives so that the 
domain with wavevector along $\vm $ remains metastable.  
Above $B_{c1}$, that metastable domain disappears.  As the field increases, the wavevectors of the stable domains rotate towards 
the orientation $\YY \perp \vm $.  In the absence of domain pinning, that rotation is complete at $B_{c2} \approx 5.5$ T.  

For each field, the dependence of energy on $\vm \cdot \xp $ can be described by a sixth order polynomial with even terms only.
Based on the polynomial fits given by the solid curves in Fig.2, we obtain the minimum energy solutions
for $\vert \vm \cdot \xp \vert $ at each field.  We plot $\vert \vm \cdot \xp \vert_{\rm min}$ versus field in Fig.3.  
Above $B_{c2}\approx 5.5$ T, $\vq $ lies perpendicular to $\vm $ and $\vert \vm \cdot \xp \vert_{\rm min} = 0$.

\begin{figure}
\includegraphics[width=8.5cm]{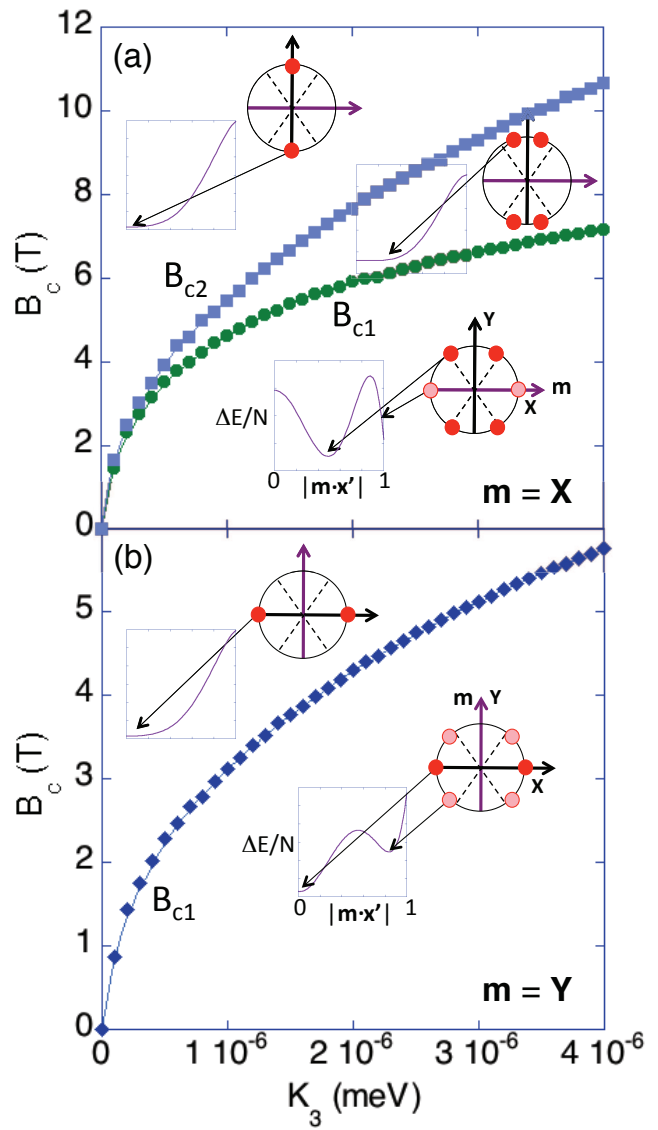}
\caption{(Color online) (a) The critical fields $B_{c1}$ and $B_{c2}$ versus three-fold anisotropy $K_3$ for $\vm =\XX $.  (b) The critical field
$B_{c1}$ versus $K_3$ for $\vm = \YY $.  Insets schematically show the
dependence of the satellites on field and their energies.}
\end{figure}

The critical fields are plotted against the three-fold anisotropy $K_3$ [\onlinecite{method}] in Fig.4(a).  
Both critical fields $B_{c1}$ and $B_{c2}$ and their difference $B_{c2}-B_{c1}$ increase quite rapidly $\sim \sqrt{K_3}$ for small $K_3$.  
We schematically sketch the dependence of the satellite peaks on magnetic field and their energies in the insets to Fig.4.

\subsection{Field perpendicular to a three-fold axis or $\vm = \YY $}

When the field lies along $\YY $, the orientation of the stable domain does not change with field, i.e.
domain 2 with $\xp = [1,0,-1] \perp \vm $ or $\vm \cdot \xp = 0$
is always stable.  But as seen in Fig.5, domains 1 and 3 with $\xp = [0,1,-1]$ and $[1,-1,0]$ or
$\vert \vm \cdot \xp \vert = \sqrt{3}/2$ are metastable for small fields and become unstable at high fields.  

\begin{figure}
\includegraphics[width=7.9cm]{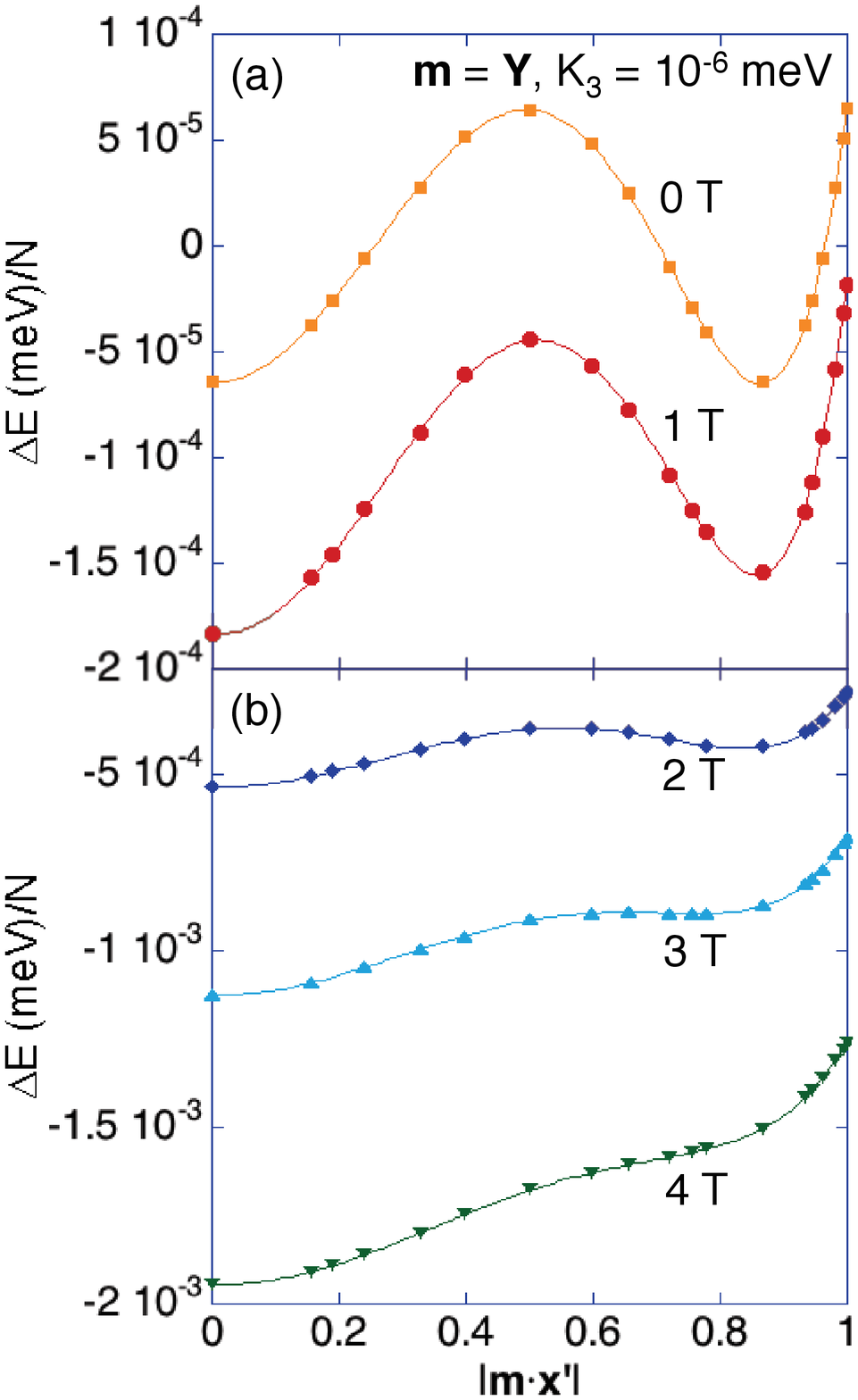}
\caption{(Color online) The energy $\Delta E/N $ versus wavevector $\vert \vm \cdot \xp \vert $ for $K_3 = 10^{-6}$ meV and field ranging from 0 to 4 T
along $\YY $.  Solid curves are fits to a sixth order polynomial in $\vm \cdot \xp $.}
\end{figure}

As in the previous subsection, $\Delta E/N$ can be fit by a sixth order polynomial in $\vm \cdot \xp $ (even
terms only).  When $K_3 = 10^{-6}$ meV, domains 1 and 3
are metastable below $B_{c1}\approx 3.1$ T.   With increasing field, the orientations $\xp $ of the metastable domains rotate slightly 
towards the three-fold axis perpendicular to $\vm $, as seen in the top curve of Fig.3.   At $B_{c1}$, $\vert \vm \cdot \xp \vert = \sqrt{2}/2$ so that 
the domain wavectors $\vq $ have rotated from $\theta = \pm \pi /6$ away from $\YY $ at zero field to $\pm \pi /4$ away from $\YY $ at $B_{c1}$
(although pinning will change this conclusion).
Because the wavevector for domain 2 is already perpendicular to $\vm $ at zero field, $B_{c2}=0$.

The dependence of $B_{c1}$ on $K_3$ is shown in Fig.4(b) [\onlinecite{method}].  Once again, $B_{c1}$ scales like $\sqrt{K_3}$ for small $K_3$.

\section{Pinning}

The effects of pinning are essential to understand the behavior of the cycloidal domains in a magnetic field.
Evidence for pinning was provided by recent SANS measurements [\onlinecite{bordacsun}].  For 
$\vm = \XX $, the wavevectors of the stable $\xp = [1,-1,0]$ and $[0,-1,1]$ domains remain unchanged up to about 5.5 T, 
above which they rotate towards $\YY \perp \vm $.  For $\vm = \YY $, the wavevectors of the metastable 
$\xp = [1,-1,0]$ and $[0,-1,1]$ domains rotate towards $\XX \perp \vm $ above about 5 T.

In a ferromagnet, domain pinning is caused by structural defects that locally change the exchange interactions and anisotropies, 
creating a complex energy landscape with 
barriers between different orientations of the magnetization $\vM = 2\mb \langle \vS_i \rangle $ [\onlinecite{jourdan07}].
No doubt, these effects are also important in cycloidal spin systems.  But
the charge redistribution determined by the cycloidal wavevector $\vq $ may be even more important.  Due to the 
strong magnetoelastic coupling in \BF  [\onlinecite{slee13, lee16b}], this charge redistribution is pinned by non-magnetic impurities.
Although the total magnetoelastic energy is independent of $\vq $ (see Appendix A), the distortions 
$\epsilon_{xx}$, $\epsilon_{yy}$, and $\epsilon_{zz}$ of the rhombohedral structure separately depend on the wavevector orientation.  
In order to rotate $\vq $, the magnetic field must drag this lattice distortion, pinned by non-magnetic impurities, through the crystal.
Of course, this charge redistribution is absent in a collinear AF.

\begin{figure}
\includegraphics[width=8cm]{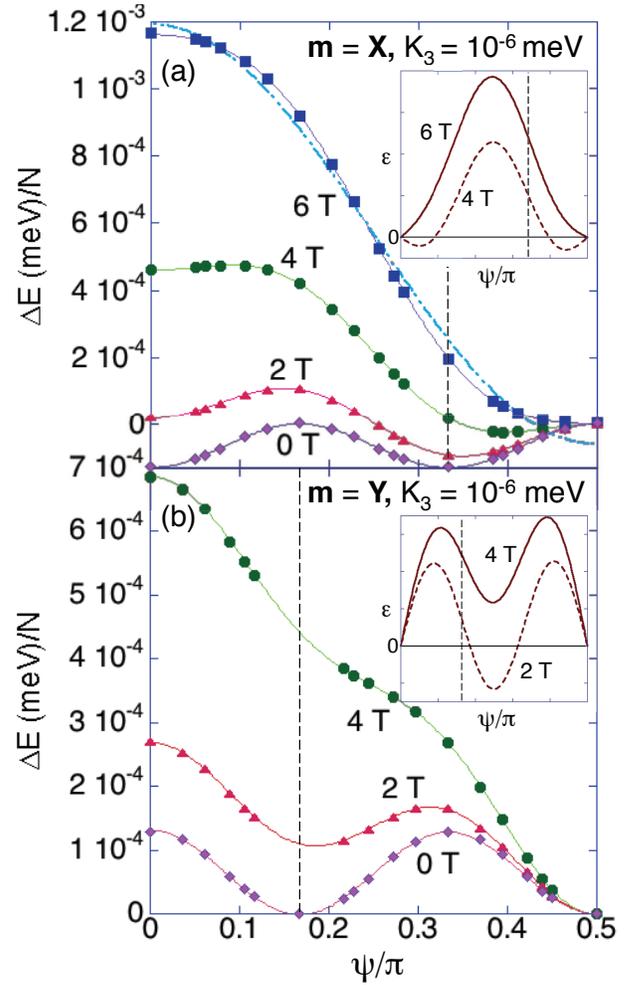}
\caption{(Color online) The energy $\Delta E/N$ versus angle $\psi = \cos^{-1} (\vm \cdot \xp )$ with $\Delta E$ set to zero at $\vm \cdot \xp = 0$
or $\psi = \pi /2$ 
for (a) $\vm = \XX $ and (b) $\vm = \YY $, both with $K_3 = 10^{-6}$ meV.  Dashed vertical lines are at $\psi = \pi /3$ and $\pi /6$, respectively.  
Insets plot the derivative $\eps $ versus $\psi $.  The dash-dot curve in (a) is a fit of the energy $\Delta E/N$ to second order in $\vm \cdot \xp = \cos \psi $.}
\end{figure}

The susceptibility $\chp $ for $\vB \parallel \yp $ or perpendicular to the plane of the cyloid
is much larger than the susceptibility $\cpa $ for $\vB $ within the cycloidal plane
[\onlinecite{leb07}].  So the induced magnetization is $\vM = \chp \vB_{\perp }$ where
$\vB_{\perp}$ is the component of $\vB $ along $\yp $.
The external field $\vB = \vB_{\perp} + \vB_{\parallel }$ plays two roles:  $\vB_{\perp }$ produces the
perpendicular magnetization $\vM $ 
and $\vB_{\parallel }$ exerts a torque $\vt = \vM \times \vB $ on $\vM $.

\vfill

\subsection{Microscopic model}

To connect these considerations with our microscopic model, Fig.6 replots $\Delta E/N$
versus $\psi = \cos^{-1} (\vm \cdot \xp )$ while setting $\Delta E/N=0$ at $\vm \cdot \xp = 0$
or $\psi  =\pi /2 $.  Using the angle definitions in Fig.7, note that 
$\psi = \theta $ for $\vm = \XX $ and $\psi = \phi =\pi/2 - \theta $ for $\vm = \YY$.
We propose that a domain is pinned until the downward 
slope $\eps = - d(\Delta E/N)/d\psi $ exceeds
the threshold $\epin >0$.  For $\vm = \XX $, $\eps $ decreases 
with $\psi $ in the neighborhood of $\psi = \pi /3$, as seen in the inset to Fig.6(a).  
As $\psi $ increases, larger fields are required to fulfill the condition $\eps > \epin $.
A similar result is found for $\vm = \YY $ near $\psi = \pi /6$, as seen in Fig.6(b).  
In both cases, $\psi $ satisfies the depinning condition $\eps = \epin $ as the field increases.

In the limit of strong pinning, the condition $\eps = \epin $ can be solved exactly.  For large fields, the anisotropy can be
ignored and $\Delta E/N = -MB\sin \psi $.  To linear order in the field, $M = \chp B \sin \psi $ so that
$\Delta E/N = -\chp B^2 \sin^2 \psi $ and $\eps = \chp B^2 \sin 2\psi $.

For $\vm = \XX $, the energy $\Delta E/N$ 
is fairly well-described by the form given above $\propto \sin^2 \psi $ at high fields, as seen by the dash-dot curve in Fig.6(a) for 6 T.
This agreement improves with increasing field.
Consequently, $\eps \propto \sin 2\psi $ is close to the form in the inset to Fig.6(a) 
near $\psi = \theta = \pi /3$ or $\phi = \pi /6$.  So for strong pinning, $\phi $ satisfies the condition
\begin{equation}
\label{bc}
\sin 2\phi = \frac{\sqrt{3}}{2} \biggl( \frac{ \Bp }{B} \biggr)^2,
\end{equation}
where the pinning field ${\Bp }^2 = 2\epin /\sqrt{3}\chp $ is defined so that $\phi = \pi /6$ when $B=\Bp $.

For $\vm = \YY $, the expression $\eps \propto \sin 2 \psi $ is 
not satisfied until fields far above $B_{c1}$.  
So the simplified expression of Eq.(\ref{bc}) 
cannot be applied when $\vm = \YY$ and $\psi = \phi = \pi /6$.  Hence,
the depinning condition $\eps = \epin $ must be solved numerically.

Nonetheless, we can still draw some qualitative conclusions.  The inset to Fig.6(b) indicates that
domains 1 and 3 become unstable when $\eps > 0$ for
$\psi =\phi = \pi /4$.  So in the absence of pinning, $\psi $ will grow from $\pi/6$ in zero field to $\pi /4$ at $B_{c1}$, in agreement with Fig.3.  
Taking pinning into account, there are two
possible ways for domains 1 and 3 to evolve with field.  If $\Bp > B_{c1}$, then the domains will disappear only after becoming depinned at
$\Bp $ with $\psi = \pi /6$.  If $\Bp < B_{c1}$, then $\psi $ will start rotating from $\pi /6$ towards $\pi /4$ above $\Bp $ and stop rotating at $B_{c1}$ 
with $\psi < \pi /4$.  So the rotation towards $\pi /4$ is not completed.

\subsection{Landau-Lifshitz-Gilbert equation}

Another way to approach pinning is through 
the Landau-Lifshitz-Gilbert (LLG) equation [\onlinecite{landau35}] for the time dependence of the magnetization:
\begin{equation}
\frac{\partial \vM }{\partial t} = -2\mb \vt + \gamma \vM \times \vt ,
\end{equation}
where $\vt = \vM \times \vB_{\rm eff}$ is the torque and $\vB_{\rm eff}$ is an effective field that includes the effect of anisotropy.
The first term in the LLG equation produces the precession of $\vM $ about $\vB_{\rm eff}$ and the second term gives the damping of 
$\vM $ as it approaches equilibrium.  So $\gamma $ is proportional to the inverse of the relaxation time of the spins.

\begin{figure}
\includegraphics[width=9cm]{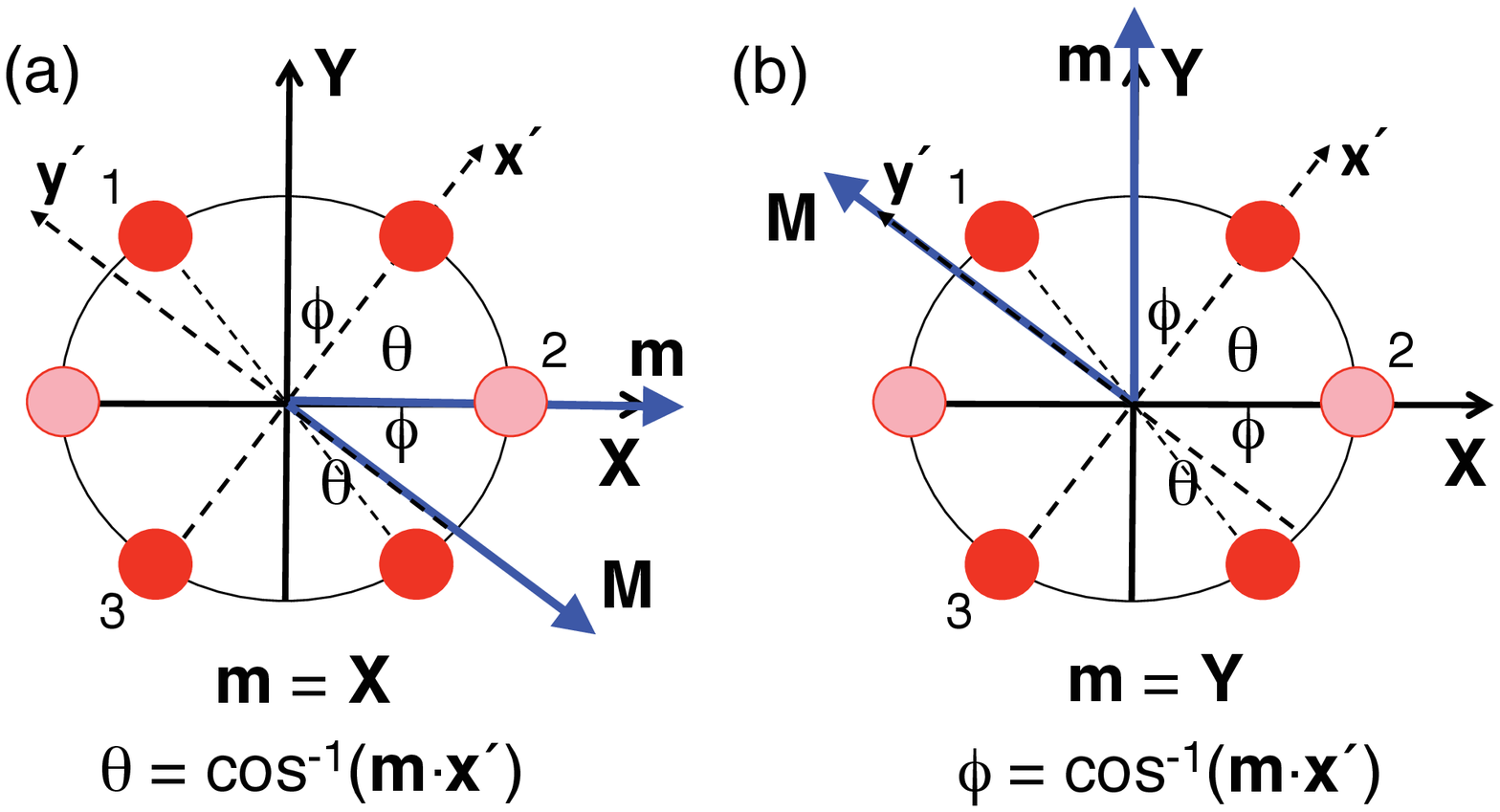}
\caption{(Color online) The magnetization and the cycloidal
$\xp $ and $\yp $ axis for domain 3 with (a) $\vm = \XX $ and (b) $\vm = \YY$.  Also shown are angles $\theta $ and $\phi = \pi/2 - \theta $.
}
\end{figure}

In the strong-pinning limit (see the discussion at the end of this section), we can neglect anisotropy and set $\vB_{\rm eff} = \vB$. 
Since $\vM $ rotates within the $\XX - \YY $ plane, it can be written
$\vM = M\bigl(\cos \varphi \,\XX +\sin \varphi \,\YY \bigr)$ so that
\begin{eqnarray}
\label{tuu}
&& \vt = \vM \times \vB =  M \Bigl\{ B_Z \bigl(\sin \varphi \,\XX - \cos \varphi \, \YY \bigr) \nonumber \\
&&+ \bigl(B_Y \cos \varphi - B_X \sin \varphi \bigr) \ZZ \Bigr\}.
\end{eqnarray}
The LLG equation then gives
\begin{eqnarray}
\frac{d\varphi }{dt} &=& -2\mb B_Z  -\gamma M \bigl(B_Y\cos \varphi -B_X\sin \varphi \bigr) \nonumber \\
& =& -2\mb B_Z - \gamma \tau_Z.
\end{eqnarray}
When $\vm = \XX $, $\varphi = -\phi$ and $M=\chp B \cos \phi $ so
\begin{equation}
\tau_Z = \frac{\chp B^2}{2} \, \sin 2 \phi  = \frac{\eps }{2}.
\end{equation}
Hence, the torque along $\ZZ $ and the energy derivative $\eps $ are simply related in the strong-pinning limit.
Ignoring the precession of $\vM $ about $\ZZ $ induced by $B_Z$, the relaxation of 
$\phi $ towards equilibrium within the $\XX - \YY $ plane is determined by $\tau_Z = \eps /2$.

Pinning in a ferromagnet is described by an effective field [\onlinecite{lemerle98, kleemann07}] 
that opposes the applied field, both along $\vM  $. 
In the strong-pinning limit of a cycloidal spin system, the external torque $\vt $ along $\ZZ $ is opposed by a pinning 
torque with maximum magnitude $\tpin $.  The conditions $\tau_Z = \tpin $ and $\eps = \epin $ are then
equivalent.  In terms of $\tpin $, the pinning field is given by ${\Bp }^2 = 4\tpin /\sqrt{3} \chp $.

For $\vm = \XX $, Eq.(\ref{bc}) is solved for $\phi $ as a function of $B/\Bp $ in Fig.8.    
As shown in the next section, Eq.(\ref{bc}) can be refined
by expanding $\chp $ in powers of $B^2 \cos^2 \phi$.  Since $\phi $ never reaches 0, $B_{c2}= \infty $.

Because the charge redistribution evaluated in Appendix A only depends on the 
direction of $\vq $, $\tau_Z$ does not depend on the interior details of the cycloid 
such as its period or higher harmonics, but only on the magnetization $\vM $ induced by $\vB_{\perp }$.   

For $\vm =\XX$, experiments [\onlinecite{bordacsun}] observe pinning when $B$ is lowered from $\Bp $ but 
not as it is raised.  This can be easily explained based on our model.   For a fixed slope $\varepsilon = \epin $,
$\Delta E /N \propto \cos^2 \theta /\sin 2\theta $ decreases with increasing $\theta \ge \pi /3$, as shown in the inset to Fig.8.  
So when $B$ is raised from $\Bp $, $\phi =\pi /2 -\theta $ relaxes to a smaller value with lower energy.
But when $B$ is lowered from $\Bp $, $\phi $ would have to take a larger value with {\it higher} energy to satisfy the condition
$\varepsilon =\epin $ or
$\tau_Z = \tpin $.  This process is energetically prohibited at low temperatures.  The pinning of 
domains with decreasing field is shown by the 
dashed lines in Fig.8.  When the field is ramped up again with this value of $\phi $, $\vq $ will only start rotating towards 
smaller values of $\phi $ when the condition given by Eq.(\ref{bc}) is reached at the solid curve.  

\begin{figure}
\includegraphics[width=9cm]{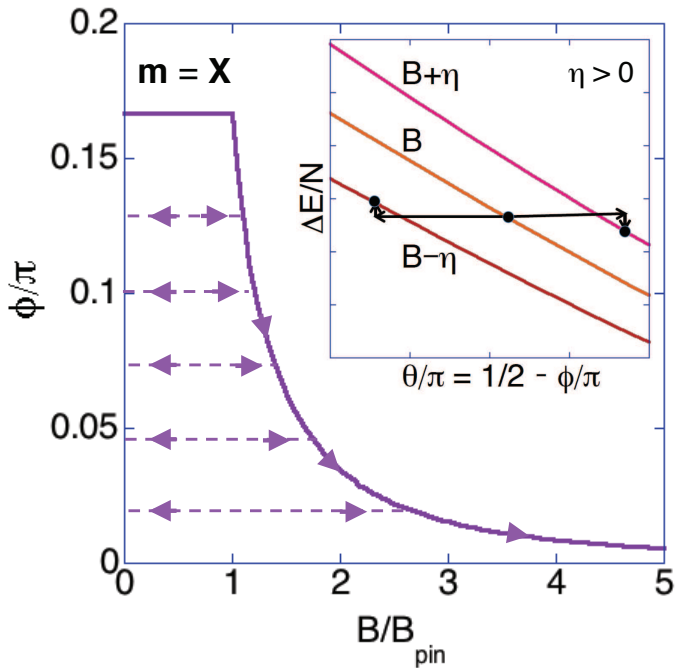}
\caption{(Color online) With $\vm = \XX $, the evolution of the rotation angle $\phi $ for domains 1 and 3 with field $B$ normalized by $\Bp $.  
The solid curve is for increasing field, the dashed lines are for decreasing or increasing field.  Inset shows that $\Delta E/N$
would have to rise with decreasing field to keep its downward slope $\varepsilon $ unchanged.  For increasing field, $\Delta E/N$
drops.}
\end{figure}

\vfill

\section{Results}

Let's use these ideas to examine the experimental results for \BP .  We separately discuss the two cases for field along $\XX $ or
$\YY $ examined in Section IV.

First take $\vm = \XX $
as in Section IV.A.  Since domain 2 with $\vq_2 \parallel \vm $ becomes unstable when
$B > B_{c1} \approx 7$ T  [\onlinecite{bordacsun}], the dependence of $B_{c1}$ on $K_3$ from 
Fig.4 implies that $K_3\approx 3.65 \times 10^{-6}$ meV.   
The pinning field $\Bp \approx 5.6$ T for domains 1 and 3 is estimated from experimental results.
Measurements by Bordacs {\em et al.} [\onlinecite{bordacsun}] confirm that the wavevectors of domains 1 and 3 never become fully perpendicular
to $\vm = \XX $ but that $\phi \rightarrow 0$ or $\theta \rightarrow \pi /2$ with increasing field.  

This model can be further refined by expanding
$\chp $ in powers of $B^2 \cos^2 \phi $.  With $b = B/\Bp $, 
\begin{equation}
\chp (b) = \chi_0 + \chi_2 b^2 \cos^2 \phi ,
\end{equation}
where $\chi_2$ is the non-linear susceptibility.  Defining $\alpha = \chi_2 /\chi_0$, we numerically solve
\begin{equation}
\label{refi}
b^2 \Bigl\{ 1 + \alpha b^2 \cos^2 \phi \Bigr\} \sin 2\phi = \frac{\sqrt{3}}{8} \,(4+3\alpha  )
\end{equation} 
with pinning field
\begin{equation}
{\Bp }^2 = \frac{4}{\sqrt{3}} \frac{\tpin }{\chi_0 + 3\chi_2 /4}.
\end{equation}
The solutions to Eq.(\ref{refi}) with $\alpha = 0.2$ and 0.4 are plotted in the dashed curves of 
Fig.9 and are in good agreement with measurements.
As mentioned above, the dependence of $\phi $ on field given by Eq.(\ref{refi}) may only be approached at large
fields if the condition for the strong-pinning limit is not met at $\Bp $.

Now take $\vm = \YY $ as in Section IV.B.  Then domain 2 with
$\vq_2 \perp \vm $ is always stable.  Unfortunately, the rotation of domains 1 and 3 cannot be treated 
in the strong-pinning limit.   In particular, $\Bp $ may differ from the earlier result for
$\vm = \XX $.
Experiments with $\vm = \YY $ indicate that domains 1 and 3 rotate by about 9 degrees before
disappearing between 6 and 7 T.  This implies that the second scenario discussed above with $B_{c1} > \Bp $ 
is applicable.  As expected, $\psi = \phi $ has only increased to about $39^\circ < 45^\circ $ at $B_{c1}$.
But for $K_3 = 3.65\times 10^{-6}$ meV, Fig.4(b) predicts that $B_{c1} \approx 5.6$ T, which is below the range
of observed values where the domains disappear.
This discrepancy can possibly be explained by a slight misalignment of the field out of the $\XX - \YY $ plane.

Since $\tpin $ depends on the concentration and 
distribution of non-magnetic impurities, it may also change in different samples.   Because the samples used in the spectroscopy [\onlinecite{nagel13}]
and SANS measurements [\onlinecite{bordacsun}] come from different sources, their pinning torques may be different as well.
Based on the relative purities of these two samples, $\Bp $ may be larger than indicated above for the sample 
used in the spectroscopy measurements.
The observed spread [\onlinecite{bordacsun}] in wavevectors $\vq_1$ and $\vq_3$ near $\YY $ for $\vm = \XX $ also suggests that the pinning torque $\tpin $ 
varies from one domain to another throughout the sample used in the SANS measurements.   Unlike $\tpin $, $B_{c1}$ is determined by the relative energies of
different domain wavevectors $\vq $ and is independent of sample quality.

\begin{figure}
\includegraphics[width=8.5cm]{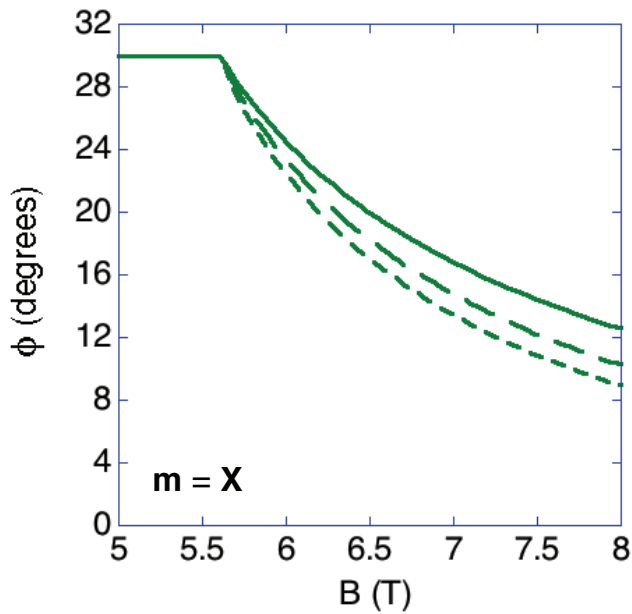}
\caption{(Color online) Model predictions for
$\Bp = 5.6$ T and $\alpha = 0$ (solid), 0.2 (long dash), or 0.4 (short dash).}
\end{figure}

The ``strong-pinning'' limit is reached when the anisotropy $K_3$ makes a negligible contribution to the energy
compared to the pinning field $\Bp $.  
Considering the contributions of the magnetic field and the anisotropy to the energy, 
the strong-pinning limit requires that $\Bp M \gg 2S^6K_3$ or 
\begin{equation}
\label{sp}
{\Bp }^2 \gg \frac{2S^6K_3}{\chp }.
\end{equation}
Using the definition of $\Bp $ in terms of the pinning torque $\tpin $, the strong-pinning limit requires that
$\tpin \gg \sqrt{3} S^6 K_3/2$.
Taking $K_3=3.65 \times 10^{-6}$ meV and $\chp $ from measurements [\onlinecite{tokunaga15}], 
these conditions require that $\Bp \gg 2.5$ T and $\tpin \gg 0.77$ $\mu $eV.

Even for small $\Bp $, the strong-pinning limit can be reached when the field is sufficiently large that 
it dominates the energy.  This condition is given by Eq.(\ref{sp}) with $B$ replacing
$\Bp $.  So independent of $\Bp $, Eq.(\ref{bc}) for the dependence of $\phi $ on field is approached when
$B \gg 2.5$ T.

\vfill

\section{Discussion}

This work resolves all of the discrepancies with earlier measurements listed in the introduction.
Not too close to the poles at $\pm \zp $, the critical field $B_{c3} > 16$ T above which the CAF phase becomes 
stable does not sensitively depend on the 
azimuthal angle $\zeta $ because $\vq $ is then nearly perpendicular to $\vm $.  
This explains the earlier discrepancy with measurements by Tokunaga {\it et al.} [\onlinecite{tokunaga10}].
For any $\zeta $, Ref.[\onlinecite{fishman13b}] then predicts that $B_{c3}$ will increase monotonically as the 
polar angle $\vartheta $ decreases from $\pi /2$ at the equator to zero at the poles.

Because it couples
to the wavevector orientation but not to the individual spins, the strain does not directly affect the spin dynamics.  However, it may be
necessary to slightly raise $K_1$ to compensate for the effect of $K_3$, which favors the
spins lying in the $\XX - \YY $ plane [\onlinecite{comp}].  Since the total
magnetoelastic energy is independent of $\vq $, it does not alter the relative energies of different wavevectors in Figs.2 and 5.  Measurements of
the lattice strain in a magnetic field along a three-fold axis like $\XX $ can test this hypothesis.

How does our estimate for $K_3$ in \BF compare with that in other materials?  
The constant $K_3$ can be estimated from the angular dependence of the basal-plane magnetization or the torque.   
For Co$_2$$Y$ ($Y$ = Ba$_2$Fe$_{12}$O$_{22}$) and Co$_2$$Z$ ($Z$ = Ba$_3$Fe$_{24}$O$_{41}$), $\tilde K_3 \equiv S^6 K_3 /V_c \approx 600$ erg/cm$^3$ and 
$1500$ erg/cm$^3$, respectively [\onlinecite{bickford60}] ($V_c$ is the volume for one magnetic ion).   For pure Co, $\tilde K_3 \approx 1.2 \times 10^5$ erg/cm$^3$
[\onlinecite{paige84}].
Anisotropy energies are much larger for the rare earths than for transition-metal oxides [\onlinecite{rhyne72}].
While $\tilde K_3 \approx 6300$ erg/cm$^3$ for Gd, it is about 1000 times higher for the heavier rare earths Tb, Dy, Ho, Er, and Tm.  
By comparison, $K_3 = 3.65 \times 10^{-6}$ meV for \BF corresponds to $\tilde K_3 = 2.4\times 10^4$ erg/cm$^3$, about 4 times larger than for Gd but smaller than for pure Co.

Our discussion of domain pinning was motivated by previous results for ferromagnets.  For a ferromagnet, thermally activated creep 
[\onlinecite{feigel89}] allows the domain walls to move even when $B < \Bp $. 
It seems likely that a similar effect in \BF allows the domains to rotate at nonzero temperature even when $\tau_Z < \tpin $
or $\eps < \epin $.

We conclude that the ``canonical" model of \BF must be augmented by three-fold anisotropy and magnetoelastic energies in 
order to explain the field evolution of a domain when $\vq $ is not perpendicular to $\vm $.  Over the past decade,
our understanding of \BF has greatly increased but so have the number of new mysteries to be solved.
At least at low temperatures, we believe that the modified Hamiltonian presented in this work can be used 
to study the manipulation of domains by magnetic and electric fields. 

Thanks to Sandor Bord\'acs, Takeshi Egami, Daniel Farkas, and Istvan K\'ezsm\'arki for helpful discussions.
Research sponsored by the U.S. Department of Energy, 
Office of Basic Energy Sciences, Materials Sciences and Engineering Division.

\vfill

\appendix

\section{Magnetoelastic Coupling}

This appendix describes the effects of the magnetoelastic coupling in \BP .   The 
magnetoelastic energy is given by
\begin{eqnarray}
&&\frac{1}{V}H_{me} =  \frac{1}{2}c_{11} \Bigl( {\epsilon_{xx}}^2 +{\epsilon_{yy}}^2 +{\epsilon_{zz}}^2 \Bigr)\nonumber \\
&&+c_{12}\Bigl( \epsilon_{xx}\epsilon_{yy} +\epsilon_{yy}\epsilon_{zz}+\epsilon_{zz}\epsilon_{xx}\Bigr)\nonumber \\
&&+\frac{g }{N} \sum_i \Bigl\{ \epsilon_{xx}{S_{ix}}^2 + \epsilon_{yy}{S_{iy}}^2 + \epsilon_{zz}{S_{iz}}^2\Bigr\},
\end{eqnarray}
where $c_{11}$ and $c_{12}$ are the elastic coupling constants, $g$ is the magnetoelastic coupling strength,
and $\epsilon_{ii}$ are the strain components.

Minimizing this energy with respect to the strain components yields
\begin{eqnarray}
\epsilon_{xx}= -\frac{g}{F} \Bigl\{ \bigl(c_{11}+c_{12}\bigr) M_{2x} - c_{12} \bigl(M_{2y} +M_{2z}\bigr) \Bigr\},\\
\epsilon_{yy}= -\frac{g}{F} \Bigl\{ \bigl(c_{11}+c_{12}\bigr) M_{2y} - c_{12} \bigl(M_{2x} + M_{2z}\bigr) \Bigr\},\\
\epsilon_{zz}= -\frac{g}{F} \Bigl\{ \bigl(c_{11}+c_{12}\bigr) M_{2z} - c_{12} \bigl(M_{2x} +M_{2y}\bigr) \Bigr\},
\end{eqnarray}
where 
\begin{equation}
M_{2\alpha } = \frac{1}{N} \sum_i \langle {S_{i\alpha }}^2 \rangle 
\end{equation}
and $F=\bigl(c_{11}+2c_{12}\bigr)\bigl(c_{11}-c_{12}\bigr)$.
In terms of these variables, the magnetoelastic energy $E_{me}=\langle H_{me}\rangle $ is given by
\begin{eqnarray}
&&\frac{1}{N}E_{me}= -g^2\frac{V}{2N}\frac{c_{11}+c_{12}}{F} \Bigl\{ {M_{2x}}^2+{M_{2y}}^2 +{M_{2z}}^2\Bigr\}\nonumber \\
&&+g^2 \frac{V}{N}\frac{c_{12}}{F}\Bigl\{ M_{2x}M_{2y}+M_{2x}M_{2z}+M_{2y}M_{2z}\Bigr\},
\end{eqnarray}
where $V/N=a^3$.

Transforming $M_{2\alpha }$ into the local reference frame of the cycloid, we find
\begin{equation}
M_{2\alpha } = \frac{{q_{\alpha}}^2}{\vert \vq \vert^2 } M_{2x' } + \frac{1}{3}M_{2z'},
\end{equation}
which uses $M_{2y'} \approx 0$.
For weak anisotropy, $M_{2x'}\approx M_{2z'} = S^2/2$ and
\begin{equation}
\frac{1}{N}E_{me}= -\frac{g^2S^4}{4F} \frac{V}{N} \bigl(3c_{11}-2c_{12}\bigr),
\end{equation}
which is independent of $\vq $.

However, the individual strain components
\begin{eqnarray}
\epsilon_{\alpha \alpha } = -\frac{gS^2}{2F} \biggl\{ \frac{{q_{\alpha }}^2 }{\vert \vq \vert^2} \bigl(c_{11}+2c_{12}\bigr) +
\frac{1}{3}\bigl(c_{11}-4c_{12}\bigr)\biggr\}\\ \nonumber
\end{eqnarray}
do depend on $\vq $.  In fact, roughly 75\% of the strain depends on the wavevector orientation.
Impurities clamp this lattice strain within the sample.
Because the magnetic field must drag this distortion while rotating the cycloidal
wavevector $\vq $, impurities pin the orientation of the wavevector at low fields.  
Note that the volume change
\begin{equation}
\frac{\Delta V}{V}=\epsilon_{xx}+\epsilon_{yy}+\epsilon_{zz}= -\frac{gS^2}{c_{11}+2c_{12}}
\end{equation}
is independent of the wavevector orientation.


\begin{thebibliography}{}
\bibitem{param} This paper uses the parameters $J_1 = -5.3$ meV, $J_2 = -0.2$ meV, $K_1 = 0.004$ meV, $D_1=0.18$ meV, and $D_2 = 0.06$ meV,
which reproduce the zero-field mode frequencies [\onlinecite{fishman13a}] scaled by $S=5/2$ rather than by $\sqrt{S(S+1)}\approx 3$.
\bibitem{other} Higher-order easy-axis anisotropy terms $-S^4 {\bar K}_2 \sum_i \cos^4 \theta_i$ and $-S^6 {\bar K}_3 \sum_i \cos^6 \theta_i$ are also consistent with 
the $R$3$c$ crystal structure of \BF but they do not produce any qualitatively new effects.  Since the easy-axis $-S^2 K_1 \sum_i \cos^2 \theta_i$ term is already
included in ${\cal H}$, we neglect those terms.
\bibitem{method} To evaluate $B_{c1}$ with $\vm = \XX $, we compare the energies of domains with wavevectors $[1,0,-1]$ ($\vm \cdot \xp = 1$) 
and $[5,1,-6]$ ($\vm \cdot \xp = 0.99$).  Below $B_{c1}$, the domain with wavevector $[1,0,-1]$ has lower energy;
above $B_{c1}$, it has higher energy.  To evaluate $B_{c2}$ with $\vm = \XX $, we compare the energies of domains with wavevectors $[-1,2,-1]$ ($\vm \cdot \xp =0$)
and $[-2,5,-3]$ ($\vm \cdot \xp = 0.11$).  Below $B_{c2}$, the domain with wavevector 
$[-2,5,-3]$ has lower energy; above $B_{c2}$, it has higher energy.  To evaluate $B_{c1}$ with $\vm = \YY $, we compare the energies of domains
with wavevectors $[-4,3,1]$ ($\vm \cdot \xp = 0.72$) and $[-3,2,1]$ ($\vm \cdot \xp = 0.65$).  Below $B_{c1}$, the domain with wavevector $[-4,3,1]$ has the lower energy;
above $B_{c1}$, it has higher energy.
\bibitem{comp} The energy differences for a spin lying in the $\XX - \YY $ plane and along $\zp $ 
are $-K_1S^2$ and $K_3 S^6$.  So it may be necessary to raise $K_1$ by $\Delta K_1 = K_3 S^4$
to compensate for the effect of $K_3$.  For $K_3 = 3.65 \times 10^{-6}$ meV,
this corresponds to a change $\Delta K_1 /K_1$ of about 3.5\%.
\end{thebibliography}
\end{document}